\pdfoutput=1
\documentclass[11pt]{article}

\usepackage{booktabs} 
\usepackage{enumerate}
\usepackage{verbatim}

\usepackage{algorithm,algorithmic,  xcolor}
\usepackage[fleqn]{amsmath}
\usepackage{times, epsfig, theorem, amssymb,graphicx, wrapfig}

\usepackage[square, numbers, sort]{natbib}

\newtheorem{definition}{Definition}

\newtheorem{lemma}{Lemma}

\renewcommand{\emph}{\textit}

\begin{document}

\title{An exploration of algorithmic discrimination in data and classification}


\author{Jixue Liu$^{a,1}$, Jiuyong Li$^{a}$, Feiyue Ye$^{b,1}$, Lin Liu$^{a}$, Thuc Duy Le$^{a}$, Ping Xiong$^{a}$\\
$^a$ University of South Australia  \hspace{3em}  $^b$ Jiangsu University of Technology, China\\
\{jixue.liu,jiuyong.li,lin.liu,thuc.le,ping.xiong\}@unisa.edu.au; yfy@jsut.edu.cn
}

\maketitle
\footnotetext[1]{supported by Grant 61472166 of National Natural Science Foundation China}

\begin{abstract}
Algorithmic discrimination is an important aspect when data is used for predictive purposes. This paper analyzes the relationships between discrimination and classification,  data set partitioning, and decision models, as well as correlation. The paper uses real world data sets to demonstrate the existence of discrimination and the independence between the discrimination of data sets and the discrimination of classification models.  
\end{abstract}



\newcommand{\jry}[1]{{\color{red}JX: #1}}
\newcommand{\calf}[1]{{{\cal #1}}}
\newcommand{\spc}[1]{{\hspace{#1ex}}}
\newcommand{\ttop}[1]{\ensuremath{\hspace{-.1ex}{{\small #1}}\hspace{-.1ex}}}
\newcommand{\tteq}{\ttop{=}}
\newcommand{\vv}[1]{\avlong#1!}
\def\avlong#1,#2!{{#1=#2}}
\newcommand{\mb}[1]{\ensuremath{\mathbf{#1}}}
\newcommand{\mbE}{\ensuremath{\mathbf{E}}}
\newcommand{\mbe}{\ensuremath{\mathbf{e}}}
\newcommand{\mbP}{\ensuremath{\mathbf{P}}}
\newcommand{\mbp}{\ensuremath{\mathbf{p}}}
\newcommand{\mbO}{\ensuremath{\mathbf{O}}}
\newcommand{\mbEO}{\ensuremath{\mathbf{EO}}}
\newcommand{\mbo}{\ensuremath{\mathbf{o}}}
\newcommand{\calP}{\ensuremath{{\cal P}}}
\newcommand{\calM}{\ensuremath{{\cal M}}}
\newcommand{\ptns}{\ensuremath{{\cal P}tns}}
\newcommand{\confto}{\ensuremath{\asymp}}
\newcommand{\bfit}[1]{\textbf{\textit{#1}}}
\newcommand{\agrp}[1]{\ensuremath{\langle \mb{#1} \rangle}}
\newcommand{\bref}[1]{(\ref{#1})}
\newcommand{\notimplies}{\ensuremath{\ \ \ \ / \hspace{-1.5em}\implies}}
\newcommand{\smallpara}[1]{\vspace{.5ex} \noindent {\bf #1}}
\newcommand{\smallsubsec}[1]{\vspace{1ex}\noindent {\bf\large #1}\\\vspace{0.5ex}}

\newenvironment{proof}{\newline\noindent{\bf Proof}\,:\ }


\section{introduction}

Discrimination means ``treating a person or particular group of people differently, especially in a worse way from the way in which you treat other people, because of their skin colour, sex, sexuality, etc'' (dictionary.cambridge.org). It can happen in law enforcement applications where  people may be unfairly treated and sentenced because of their races and religions, in bank loan applications where people may not get a loan because they live in a suburb with a lower economic status. 

The law does not allow discrimination to happen. That is, decisions should not be made based on people's sex, skin colour, religion etc., called protected attributes of individuals. However discrimination is still a concern in the real world. In car insurance, the insurance company required  people in a specific suburb to pay higher premium than those in other suburbs with the reason that the suburb has a higher claim rate.  If most dwellers of the suburb are of a certain race, the higher premium forms actually discrimination to the people of this race. Boatwright (2017) reported three real world lending discrimination cases \cite{threelendingDsc}. 

Discrimination does exist in the data collected from real world applications. We will demonstrate this fact in the experiment section of this paper. 

The level of discrimination is context-based. Contexts are defined by \emph{explanatory attributes}. Assume that an investigation is on whether female employees are less paid. Then, the explanatory attributes like profession and position held by the employees matter. A fair comparison between the payments of female employees and male employees must assume that the employees hold same profession and same position. Without a context, the comparison may between the income of a group of male CEOs with the income of a group of male kitchen hands. We use an example to show how discrimination levels are affected by different contexts.  

\vspace{0.3ex}
\noindent {\bf Example 1} 
Table \ref{ex-1} shows the tuple frequencies of a data set where $D$ standing for $Income$ (1=high, 0=low) is the outcome/class/target attribute, G standing for Gender and is a protected attribute, and a number in the middle of the tables is the number of tuples in $r$ having the same $(D, G)$ value. The discrimination score for females in $r$ in Part (a) is 0 where the score is defined as the percentage of high income females taking away the percentage of high income males $\delta(r,G)=P(D\tteq 1| G\tteq F)-P(D\tteq 1| G\tteq M)=\frac{10}{10+40}-\frac{15}{15+60}=0$  \cite{Calders2010}.  

For the same application, if a context is defined by an explanatory variable $S$ (standing for employment sector, S=private or public), the data set $r$ in (a) is divided into two subsets $r_1$ and $r_2$, and the tuple frequencies for each subset are shown in Parts b.1 and b.2 respectively.  The discrimination scores for the corresponding employment sectors are 0.22 and -0.24 respectively.  The scores in (a) and (b) conclude that contexts change discrimination levels.  $\Box$

\begin{table}  \begin{center} \small
\caption{Data split \label{ex-1}}
\begin{tabular}{|l|l|l|}
  \multicolumn{3}{l}{ (a) $r$ } \\
\hline
	 D & Female  & Male   \\ \hline
	1 &   10    &   15  \\  \hline  
	0 &   40   &  60  \\  \hline
	\multicolumn{3}{l}{$\delta(r,G) = 0 $}
\end{tabular}  \quad  \quad
\begin{tabular}{|l|l|l|}
\multicolumn{3}{l}{ (b.1) $r_1$ for Sector=1}\\
\hline
	D  & Female & Male  \\ \hline
	1  &   9     &  3  \\  \hline  
	0  &   20  &  30   \\  \hline
	\multicolumn{3}{l}{$\delta(r_1,G) = 0.22 $}
\end{tabular} 
\begin{tabular}{|l|l|l|}
\multicolumn{3}{l}{ (b.2) $r_2$  for Sector=0}\\
\hline
	D  & Female & Male  \\ \hline
	1  &   1      & 12  \\  \hline  
	0  &   20      &  30   \\  \hline
	\multicolumn{3}{l}{$\delta(r_2,G) = -0.24 $}
\end{tabular}
\end{center}
\end{table}

Classification models built from a data set may become discriminatory too. The discrimination of a model is the discrimination of the predictions that the model produces.  Consider data in Table \ref{tab:mod-ex}(a) where each record is for an individual. The column $M$ (1=high,0=low) is the performance measurement of individuals. Without an assumed context, this data set has no discrimination. If a predictive model is build on $r$, the model would be $\hat{D}=M$. We note that gender $G$ is protected and cannot be used in the model. When the model is applied to predict the outcome $\hat{D}$ for the individuals, the predicted data set is $\hat{r}$ in Part (b). It is easy to see that the discrimination score for $\hat{r}$ is 0.5. This confirms that classification models can be discriminatory even if the training data is fair. This type of models are said to be  {\it algorithmic discriminatory}.

\begin{table}[h]  \begin{center} \small
\caption{Model changes discrimination.  \label{tab:mod-ex}}
\begin{tabular}{|l|l|l|}
\multicolumn{3}{l}{ (a)  Data $r$}\\
\hline
	D  & G & M   \\ \hline
	1  &   F  &  1  \\  \hline  
	0  &   F      &  0   \\  \hline
	1  &   M     &  0   \\  \hline
	0  &   M     &  0   \\  \hline
\end{tabular}  \quad
\begin{tabular}{|l|l|l|}
\multicolumn{3}{l}{ (b) Prediction $\hat{r}$} \\
\hline
	$\hat{D}$  & G & M   \\ \hline
	1  &   F  &  1  \\  \hline  
	0  &   F      &  0   \\  \hline
	0  &   M     &  0   \\  \hline
	0  &   M     &  0   \\  \hline
\end{tabular}  \\
Model from training data is $\hat{D}=M$.\\
$\delta(r,G)=0$ but $\delta(\hat{r},G)=0.5$
\end{center}
\end{table}

Algorithmic discrimination becomes more a concern in automated decision making systems as machine learning and data analytics are used more and more in real applications. In loan approval cases, it is a normal practice that an automated system automatically scores and makes decisions on customer's applications. Fairness of automated decision systems is important to service receivers and to a harmonic society.  

Research work on algorithmic discrimination are from two areas: detection and removal. This paper, instead of continuing on these two directions,  goes to the fundamental side of these two areas. It aims to explore deeper understanding of discrimination related issues. It analyzes some properties of algorithmic discrimination and explores the level of discrimination in real world data sets and how classification models affect the discrimination levels. More specifically, the paper will show results on the following points.
\vspace{-.5ex}
\begin{itemize}  \topsep=0mm  \parsep=0mm \itemsep=0mm \parskip=0cm 
 \item  sources of discrimination in classification models and the relationship between fair data and the discrimination of classifiers,  
 \item  discrimination when data sets are merged and split, 
 \item  the relationship between discrimination and explanatory variables in a decision tree, 
 \item  the interaction between discrimination and correlation, 
 \item  experimental exploration of the existence of discrimination in real world data sets and classification models. 
\end{itemize}

The organization of the paper is the following. Section 2 presents a discrimination score and group discrimination. Section 3 presents our analysis of properties of discrimination. In Section 4, we show the results from experimental exploration of real world data sets.  Section 5 presents related work and the final section concludes the paper.

\section{Definitions and Problem}
In this section, we define basic notation, and present a measurement for algorithmic discrimination. 

Let $r$ be a data set on a schema $\mb{R}$ of binary attributes/variables. The attributes in $\mb{R}$ are of four types: an outcome/target/class attribute $D$, some protected attributes $\mbP$, some explanatory attributes $\mbE$, and other attributes $\mb{O}$: $\mb{R}=\{D\}\cup \mbP \cup \mbE \cup \mb{O}$. For the outcome attribute $D$, $D\tteq 1$ means a favorite outcome like $Income\tteq High$ or $Application\tteq Successful$ that an individual prefers to receive. For a protected attribute $P\in \mbP$, $P\tteq 1$ (e.g. Sex\tteq Female or Race\tteq Black) means a group of individuals who are protected by the law not to to be discriminated. The explanatory attributes  $\mb{E}$ explain why some people receive favorite outcomes more or less frequently than others or identify such people. For example, profession and education are often taken as explanatory attributes. Surgeon is a profession and people who are surgeons are mostly high income earners, while kitchen hand is another profession and people who are kitchen hands are low income earners. Explanatory attributes are often used in selection criteria of employment. The `other' attributes $\mb{O}$ are that attributes not in $(\{D\}\cup \mbP \cup \mbE)$. Some of these attributes may be correlated with the outcome variable $D$ and are used in classifiers for predictions, and some, called red-line attributes, may be correlated to some protected attributes which makes discrimination analysis challenging. 

An  {\it E-group} (or stratum of $\mb{E}$) is a subset $e$ of all the tuples having the same attribute-value pairs $\mb{E}\tteq \mb{e}$ on all explanatory attributes $\mb{E}$ in the data set $r$. $\mb{E}\tteq \mb{e}$, or equivalently  $(E_1\tteq e_1, ..., E_k\tteq e_k)$, is called the signature of the group and is denoted by $e.sig$. The concept of an E-group is fundamental in our discrimination definition. All groups of $\mb{E}$ in $r$ are denoted by $stra(\mb{E},r)$. In the case where $\mb{E}$ is empty, the whole data set $r$ is an E-group. 

Similar to an E-group, we also use $P\tteq 1$ group to mean all the tuples with $P\tteq 1$ in $r$.  

\subsubsection*{Discrimination score}
We employ the well cited discrimination score defined in  \cite{Calders2010} and score is $Pr(D\tteq 1|P\tteq 1)   -Pr(D\tteq 1|P\tteq 0)$. Other terms for this score are  risk difference \cite{Ruggieri2014} and selection lift \cite{Pedreschi2009}. In the case where $D$ is income and $P$ is gender, the score reflects the probability difference of high income earners caused by gender difference. Considering E-groups, the score for each E-group is: 
\begin{align}
\delta(P,\mb{e})  =  Pr(D\tteq 1|P\tteq 1,\mbE\tteq \mb{e}) -Pr(D\tteq 1|P\tteq 0,\mbE\tteq \mb{e}) \label{eq:ce-e}
\end{align}

The discrimination score of $P\in \mb{P}$ in {\it data set $r$} is the E-group size weighted average \cite{Zhang2016,Li-bigdata2017}:
\begin{align}
   \delta(P, r) & =  \underset{e\in stra(\mb{E},r)} {\sum} \delta(P,e)*\frac{|e|}{|r|} \label{eq:ce-r}
\end{align}
where $e$ is each $\mb{E}$ stratum and is also overloaded to represent the signature $\mb{E}\tteq \mb{e}$ of $e$ in Formula \bref{eq:ce-e}. Obviously the following Lemma is true because of the average.
\begin{lemma}\label{lemm-r-e}
$|\delta(P, r)|\le max_{e\in stra(\mb{E},r)}\{|\delta(P, e)|\}$. That is, for a given protected attribute $P$, the score of the data set is less than or equal to the maximal score of E-groups.  
\end{lemma}

A data set has multiple protected variables. The discrimination score of a data set is:
\begin{align}
   \delta(r) & =  \underset{P\in \mbP} {max} \ \delta(P, r) \label{eq:ce-PR}
\end{align}

\begin{definition}[Discrimination]
Given a data set $r$ and a user-defined discrimination score threshold $\alpha$, 
\begin{itemize}
 \item a group of a protected attribute $P$ is \bfit{group-discriminated} in the explanatory group $\mb{E}\tteq \mb{e}$  if \  $|\delta(P, \mb{e})|> \alpha$;  

\item  a group of a protected attribute $P$ is \bfit{globally discriminated} if \ $|\delta(P, r)|> \alpha$, and  

\item an E-group $e$ is \bfit{discriminatory} if $\exists P\in \mb{P}: |\delta(P, r)|> \alpha$. 

\item the data set $r$ is \bfit{discriminatory} if $|\delta(r)|>\alpha$. $r$ is \bfit{discrimination-safe} if $|\delta(r)|\le\alpha$. $r$ is \bfit{discrimination-free} if $\delta(r)=0$. 
\end{itemize}
\end{definition}

Consider a classification model $M$ and a data set $r$ on schema $\{D\}\cup \mbP \cup \mbE \cup \mbO$. When $M$ is applied to $r$, a new outcome $d$ is predicted for each tuple $t \in r$. We use $d$ to replace the value $t[D]$ in column $D$ of $t$,  add hat to the attribute $D$, as $\hat{D}$, to reflect such a change, and denote the updated data set by $\hat{r}$ which is now on the schema $\{\hat{D}\}\cup \mbP \cup \mbE \cup \mbO$. $\hat{r}$ is called the \emph{ predicted data set} of $r$ by $M$.

\begin{definition}[Discrimination of a model]
Given a data set $r$, a classification model $M$, its predicted data set $\hat{r}$,  and a user-defined discrimination score threshold $\alpha$, $M$ is discriminatory if $\hat{r}$ is discriminatory with regard to the outcome $\hat{D}$ (instead of $D$). 
\end{definition}

\section{Properties of discrimination }

In this section, after introducing the notation to calculate Formula \bref{eq:ce-e}, we present our results of discrimination properties related to the building of classifiers. Our results are on the basis of an E-group, but they are also true for the whole data set as a data set is a special case of an E-group. 

To calculate the probabilities in Formula \bref{eq:ce-e}, we partition an E-group $e$ into divisions, called \bfit{DP-divisions}, based on the values of the outcome variable $D$ and a protected attribute $P$. Each division is a subset of all tuples with the same $D$ and $P$ value in $e$.  The concept of a division is the same as a stratum or a group, but it is used here to make the terminology distinct. $e$ thus has four DP-divisions because $D$ and $P$ are all binary. The tuple count of each division is denoted by $f_{ij}$ where the first subscript $i$ means the $D$ value and the second  subscript $j$ represents the $P$ value. For example, $f_{11}$ means the tuple count in the division $(D=1,P=1)$. The symbols denoting the counts are defined in Table \ref{tb-count-DP}, called the \bfit{counts table}. 

\begin{table}[h] \small\begin{center}
\caption{Tuple counts of DP-divisions		 \label{tb-count-DP}}
\begin{tabular}{|c |c c|} 
  \multicolumn{3}{c}{(a) tuple counts  of training data} \\ 
  \hline
		   & $P$=1 & $P$=0  \\ \hline
	$D$=1  &   $f_{11}$    &  $f_{10} $       \\  
	$D$=0  &   $f_{01}$    &  $f_{00}$     \\  \hline
\end{tabular} \qquad \qquad
\begin{tabular}{|c | c |c c|} 
  \multicolumn{4}{c}{(b) tuple counts  of predictions} \\ 
  \hline
		  &        pred.                & $P$=1 & $P$=0  \\ \hline
	$D$=1  & $\hat{D}$=1  &   $f^r_{11}$    &  $f^r_{10} $         \\  
	           & $\hat{D}$=0  &   $f^w_{11}$    &  $f^w_{10} $           \\ \hline 
	$D$=0  & $\hat{D}$=0  &   $f^r_{01}$    &  $f^r_{00}$   \\  
	           & $\hat{D}$=1  &   $f^w_{01}$    &  $f^w_{00}$    \\  \hline
\end{tabular} 
\end{center}
\end{table}

With the notation of the tuple counts of DP-divisions, Formula \bref{eq:ce-e} can be represented in Formula \bref{eq-count-dc}. Obviously each fraction in the formula is bounded by 1 and as a result, $\delta(P,e)$ is bounded to $[-1,1]$. 
\begin{align}
 & \delta(P,e) = \frac{f_{11}}{f_{11}+f_{01}}-\frac{f_{10}}{f_{10}+f_{00}}	\label{eq-count-dc} 
  \end{align}

In the special cases where there is no tuple in the contrast divisions for the protected attributes, i.e., $f_{11}=f_{01}=0$ or $f_{10}=f_{00}=0$, the discussion of discrimination in this case is not meaningful and no discrimination is possible. Then, $\delta(P,e)$ is defined to be 0.

When a classification model is applied to $e$, the model draws a decision boundary through the space defined by $(D,P)$. This boundary then splits the counts in Table \ref{tb-count-DP}(a) into the counts in Table \ref{tb-count-DP}(b) with the following constraints: $f_{ij}=f^r_{ij}+f^w_{ij}$ ($i,j=0,1$) where $f^r$ is the count of correct predictions and $f^w$ is the count of wrong predictions. We note that the decision boundary does not change the value of the values of the protected attributes, so correct predictions and wrong predictions are within the same P-group.  The discrimination score $\hat{\delta}(P,e)$ of  the model is calculated based on the predicted values $\hat{D}$ as the following.

\begin{align}
  \hat{\delta}(P,e) & = \frac{f^r_{11}+f^w_{01}}{f^r_{11}+f^w_{01}+f^w_{11}+f^r_{01}}-\frac{f^r_{10}+f^w_{00}}{f^r_{10}+f^w_{00}+f^w_{10}+f^r_{00}} \notag\\
  & =\frac{f^r_{11}+f^w_{01}}{f_{11}+f_{01}}-\frac{f^r_{10}+f^w_{00}}{f_{10}+f_{00}}	\label{eq-count-pred} 
  \end{align}

\subsection{Where is discrimination from?}
\begin{figure}[h]
\center      
\includegraphics[scale=0.9]{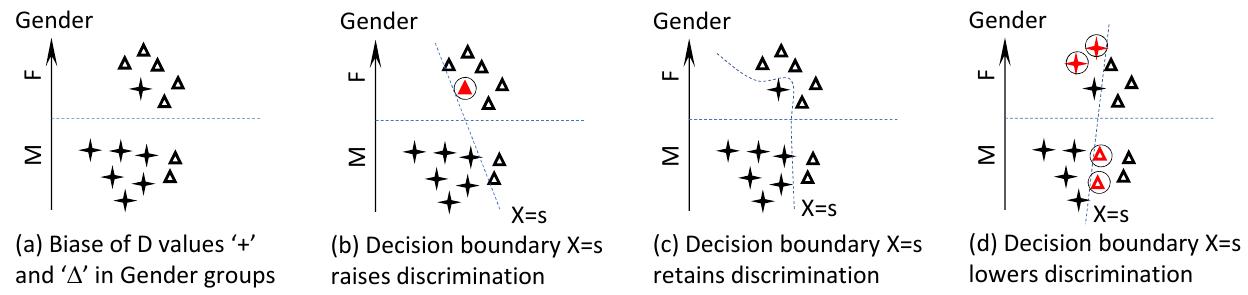}
\caption{Where is discrimination from? }
\label{fig:where-dsc}
\end{figure}

\emph{Discrimination of the (training) data is from the uneven distribution of the preferable outcome ($D$=1) in a P-group in contrast to the non-P-group.} As shown in Figure \ref{fig:where-dsc}(a) where the protected attribute $P$ is gender and the outcome values are `+' are `$\triangle$', the fraction of `+' among females is much less than the fact of `+' among males. In terms of Formula \bref{eq-count-dc}, $\frac{f_{11}}{f_{11}+f_{01}}-\frac{f_{10}}{f_{10}+f_{00}}=\frac{1}{6}-\frac{6}{8}=-0.58$ and so discrimination exists on the gender attribute. 

In the case where  a classification model is learnt from the training data, the discrimination of the model is the discrimination of predictions. Following Formula \bref{eq-count-pred}, we have the following results in Lemma \ref{lemma-perfect-model}. 

\begin{lemma}\label{lemma-perfect-model} Given a data set and a classification model trained from the data, 
\vspace{-.5ex}
\begin{itemize}
\topsep=0mm  \parsep=0mm \itemsep=0mm \parskip=0cm \itemindent=1em
\item[(1)] if the model is perfect (does not make any error in predictions), the discrimination of the model is the same as the discrimination of the training data.  If the model is perfect and the training data is discrimination-free, the model is discrimination-free. 
\item[(2)] If the model is not perfect, the discrimination of the model and the discrimination of the training data is independent. 
\end{itemize}
\end{lemma}

\paragraph{Proof} 
Item (1)  is correct because when the model is perfect $f^w_{01}=0$, $f^r_{11}=f_{11}$, $f^w_{00}=0$ and $f^r_{10}=f_{10}$, so Formula \bref{eq-count-pred}  becomes Formula \bref{eq-count-dc}. 

Item (2) is correct because if the model is not perfect, some $f^w\not=0$. Consider a case where  $f^w_{01}=0$,   $f^w_{00} =f_{00}$, $f^r_{11}=f_{11}$, and $f^r_{10}=0$. In this case, $\hat{\delta}(P,e)=1$ and this is irrelevant to the values of $f_{11}$, $f_{01}$, $f_{10}$, $f_{00}$, and consequently irrelevant to $\delta(P,e)$. \quad $\Box$

Figure \ref{fig:where-dsc}(b-d) illustrate the lemma. In Part (b), one `+' in the female group was wrongly predicted, leading the score of the predicted data to be $\frac{0}{6}-\frac{6}{8}=-0.75$, an increase in absolute value compared to that of Part (a). 

In Part (c), the predicted outcomes are identical to the original outcomes, and the discrimination score of the predictions are the same as that of the original data in Part (a).  In Part (d), the decision boundary X=s leads to a score of  $\frac{3}{6}-\frac{4}{8}=0$ in the predictions. 

This analysis shows that when a data set is discriminatory, the models learnt from it may or may not be discriminatory and this is irrelevant to the error rate levels. A small error rate can lead to large discrimination while a large error rate may lead to 0 discrimination. 

Next we show that \emph{the classifiers trained from non-discriminatory data may be discriminatory}. This seems against intuition, but it is correct. The reason for a classifier trained on non-discriminatory data not to be safe is because the input variables used in the classifier for prediction may be independent to the protected variables. When the values of the outcome variable are fair to the protected and the non-protected groups, the predicted outcomes may break this fairness, making the model unfair.  

\begin{figure}[h]
\center      
\includegraphics[scale=0.9]{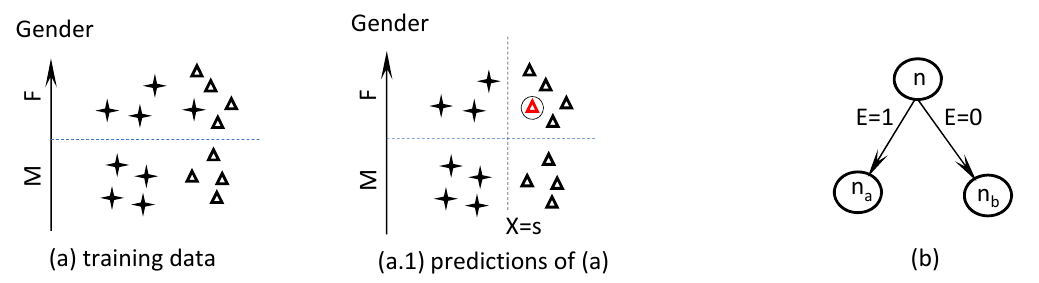}
\caption{Discrimination of model predictions }
\label{fig:mod2dsc}
\end{figure}

Consider Figure \ref{fig:mod2dsc}(a) where the outcomes are fair with regard to the Gender values and 0.5 fraction of points are getting the `+' outcome in both M and F groups, and the discrimination score is 0. 

Assume that a learning process learns the decision boundary $X=s$ and this boundary maximizes the accuracy of the model.  The predictions of the model is shown in Figure \ref{fig:mod2dsc}(a.1) and the predictions contain an error indicated by the circle.  The predictions are now discriminatory as Females are getting less number of `+' label. 

Finally, redline attributes \cite{Kamrian2012-quanti} in a model affect the discrimination of the model but this effect relates to the discrimination of the training data and the error distribution along decision boundary. Redline attributes are those correlated with a protected attribute. If this correlation is high and the distribution of the preferable outcome in the protected groups are even in the training data,  redline attributes would not be good predictors and their participation in the model is less possible. If the training data itself is discriminatory, using redline attributes in the model has a high chance to make better predictions, generate discrimination in the predictions, and make the model discriminatory. Interestingly as shown in Lemma \ref{lemma-perfect-model}, the model discrimination is also affected by the error distribution along the decision boundary.  In some cases, the decision boundary may reduce the impact of redline attributes on the model discrimination.

\subsection{Discrimination in subsets of an E-group}
In decision tree learning algorithms, a data set is split into subsets and the subsets are moved to the child nodes. We like to know how discrimination changes as a data set is split and as subsets are combined into one set. 

Consider an E-group $e$ and its sub sets $e'$ and $e''$ where $e=e'\cup e''$. 
Discrimination of $e$, $e'$ and $e''$ follows the Simpson Paradox \cite{simpson-paradox}. The lemma below shows this paradox. It indicates that even if the data set $e$ is not discriminatory, its subsets may still be discriminatory, and if every subset is non-discriminatory, the combined data set from the subsets may still be discriminatory. 

\begin{lemma}\label{lemma-simpson}
Assume that data sets $e$, $e'$ and $e''$ belong to the same E-group such that $e=e'\cup e''$. Then, the discrimination in $e$ with regard to a protected attribute $P$ is not guaranteed by the discrimination of $e'$ and $e''$ or vice versa. That is, for a user specified discrimination score threshold $\alpha$ \\
(1) $|\delta(P,e)|\le \alpha   \notimplies  |\delta(P,e')|\le \alpha$ and $|\delta(P,e'')|\le \alpha$, and   \\
(2) $|\delta(P,e')|\le \alpha$ and $|\delta(P,e'')|\le \alpha  \notimplies   |\delta(P,e)|\le \alpha$. 
\end{lemma}

We use the following proof to show a method to construct clusters of cases for the lemma.  
	\vspace*{-2ex}		\paragraph{Proof} 
\begin{enumerate}[(1)]
\item  Proof of (1). The following example shows that $e$ has a discrimination score of 0, but after partition, its subsets $e'$ and $e''$ both have maximal discrimination scores. In the example, $K$ is a non-zero positive integer and the tuple counts meet the requirements of $e=e'\cup e''$.
\[\left[ \begin{array}{ll} \multicolumn{2}{l}{e: \ \delta(e)=0} \\
                                     f_{11}=K  &  f_{10}=K  \\  
                                     f_{01}=K  & f_{00}=K \end{array}\right] \quad 
  \left[ \begin{array}{ll} \multicolumn{2}{l}{e': \ \delta(e')=1} \\
                                    f'_{11}=K  &  f'_{10}=0 \\  
                                    f'_{01}=0  & f'_{00}=K \end{array} \right] \quad         
  \left[ \begin{array}{ll} \multicolumn{2}{l}{e'': \ \delta(e'')=-1} \\
					    f''_{11}=0  &  f''_{10}=K \\ 
					    f''_{01}=K  & f''_{00}=0 \end{array} \right] 
  \] 

\item
Proof of (2).  
We show that there exist cases where the discrimination scores of subsets $e'$ and $e''$ are less than the threshold $\alpha$, but after merging $e'\cup e'' => e$, the discrimination score of $e$ is more than the threshold $\alpha$. Let $m,K$ be non-zero positive integers, and $\alpha'$ be a number so that $\alpha'<\alpha$. We choose the tuple counts to be:
\[\left[ \begin{array}{ll} \multicolumn{2}{l}{e': \quad \delta(e')=0 < \alpha} \\
                                    f'_{11}=2\alpha'mK  &  f'_{10}= \alpha' mK\\  
                                    f'_{01}=2K-2\alpha'mK  & f'_{00}=K-\alpha' mK \end{array} \right] \]
 \[ \left[ \begin{array}{ll} \multicolumn{2}{l}{e'': \quad \delta(e'')=\alpha' < \alpha} \\
                                    f''_{11}=\alpha' K  &  f''_{10}=\alpha' K \\
                                     f''_{01}=K-\alpha' K  & f''_{00}=2K-\alpha' K \end{array} \right] \]
\[  \left[ \begin{array}{ll} \multicolumn{2}{l}{e=e'\cup e'': } \\
				    f_{11}=2\alpha' mK +\alpha' K  &  f_{10}=\alpha' mK +\alpha' K\\  
				    f_{01}=3K-2\alpha' mK-\alpha' K   & f_{00}=3K- \alpha' mK-\alpha' K \end{array} \right] 
  \] 
Then, $\delta(e) = \frac{m\alpha'}{3}$. If $m>\frac{3\alpha}{\alpha'}$, then $\delta(e)>\alpha$. 

Item (2) is approved. 
\end{enumerate}
 $\Box$

The importance of Lemma \ref{lemma-simpson} is that the consideration of discrimination in subsets of an E-group does not lead to correct discrimination guarantee. Discrimination has to be analyzed against the whole of an E-group.

\subsection{Discrimination of a decision tree}
Following Lemma \ref{lemma-simpson}, we know that discrimination cannot be analyzed at subset level of an E-group. On the other hand, the discrimination of a data set can be averaged over the scores from the scores of multiple E-groups of the data set (Formula \bref{eq:ce-r}). By Lemma \ref{lemm-r-e}, if every E-group is non-discriminatory, the whole data set is non-discriminatory. 

Now we use this to analyze the discrimination score of the predictions of a decision tree. 

A leaf node of a decision tree makes predictions of a fixed label decided by majority voting during the training phase. Any tuple direct to this node by the decision path of the tree will get this label. So all the {\bf predictions} out of the leaf node has only one outcome: either $D=1$ or $D=0$. As a result, the discrimination score of the predictions out of this leaf node is 0 (see paragraph following Formula \bref{eq-count-dc}). 

Consider  Figure \ref{fig:mod2dsc}(b) with two leaf nodes of a parent node $n$  where the decision/splitting attribute $E$ is explanatory. The discrimination score of the predictions of every leaf node is 0 as explained above. As the splitting is by an explanatory attribute, the discrimination score of the predictions out of the parent node is the average of the scores on the leaf nodes and is 0.  From this, we draw the conclusion in Lemma \ref{lemma-dsc-dtree}. 

\begin{lemma}\label{lemma-dsc-dtree}
In a decision tree, if the splitting attribute of every internal node is explanatory, the predictions out of the whole decision tree are discrimination-free. 
\end{lemma}

We note that in general, the result in the above lemma may not be right. However, because all decision attributes are explanatory, the uneven distribution of the favorite outcome among different leaf nodes are `explained' by the explanatory variables in the internal nodes. The explanation  leads to 0 score in the whole tree.   This result extends to the general case. 

\begin{lemma}\label{lemma-non-dsc-E}
If a classification model ${\cal M}$ uses explanatory variables $\mbE$ as input variables,  ${\cal M}$ is non-discriminatory.
\end{lemma}

Lemma \ref{lemma-non-dsc-E} is correct because $\hat{D}={\cal M}(\mbE)$. All tuples of an E-group will have the same prediction and consequently the E-group is discrimination-free. The discrimination score averaged over all E-groups is 0. So the lemma is correct.

\subsection{Discrimination and correlation between $D$ and $\mbP$}
Correlation between a protected attribute $P$ and the outcome attribute $D$, denoted by $corr(P,D)$, is critical to discrimination. If $corr(P,D)$ is high, i.e., $P$=1 implies $D$=1, any classification model that has high accuracy will produce high level of discrimination. In this section, we analyze how discrimination is related to $corr(P,D)$.  We show that although there is a link between discrimination and correlation of $D$ and $P$, the relationship is not monotone.

The correlation between the outcome $D$ and a protected attribute $P$ can be measured by odds-ratio $oz=\frac{f_{11}*f_{00}}{f_{10}*f_{01}}$.  $oz=1$ means that there is no correlation between $D$ and $P$. $oz$ values further away from 1 indicate strong correlation. For example, if all females are high income earners ($f_{01}=0$) and all males are low income earners ($f_{10}=0$),  $D$ and $P$ are extremely correlated and the $oz$ value is infinite. 

If $oz=\frac{0}{0}$, we define $oz=\frac{0}{0}:=1$ to mean not correlated. If $f_{11}=f_{01}=0$, the $P$=1 group does not have any people, correlation does not make sense. If If $f_{11}=f_{10}=0$, no tuple has $D$=1 outcome and the case does not make sense either. The case where $f_{00}=0$ is symmetric. 

Odds-ratio and discrimination are related. As an example, assume that if all counts, except for $f_{11}$, are fixed in $oz$. In the discrimination, assume that $\frac{f_{11}}{f_{11}+f_{01}}>\frac{f_{10}}{f_{10}+f_{00}}$. Then as $f_{11}$ increases, both $oz$ and $\delta(P,e)$ increases. This trend is NOT true if other counts are allowed to change.   	  

Interestingly, the following lemma shows that the interaction  of odds-ratio and discrimination is not simple.  

\begin{lemma} \label{lemma-correla} Given an E-group $e$ and its DV-division tuple counts,
\vspace{-0.5ex}   
\begin{enumerate}[(1)]      \topsep=0mm  \parsep=0mm \itemsep=0mm \parskip=0cm \itemindent=1em
\item   $e$ is discrimination-free ($\delta=0$) if and only if  $D$ and $P$ are independent. 
\item  Less correlation does not mean less discrimination. 
\end{enumerate}
\end{lemma}

 	\vspace*{-2ex}		\paragraph{Proof} 
\begin{enumerate}[(1)]
\item 
We transform Formula \bref{eq-count-dc} into $\delta(P,e) = \frac{1}{1+\frac{f_{01}}{f_{11}}}-\frac{1}{1+\frac{f_{00}}{f_{10}}}$. When $\delta=0$, then $\frac{f_{01}}{f_{11}}=\frac{f_{00}}{f_{10}}$ and vice versa. $\frac{f_{01}}{f_{11}}=\frac{f_{00}}{f_{10}}$ means $oz=\frac{f_{01}*f_{10}}{f_{11}*f_{00}}=1$.  

The extreme cases for $\delta=0$ and the extreme cases for $oz=1$ are corresponding. So the item is proved. 

\item
We assume that we have two E-groups $e_1$ and $e_2$. Their tuple counts are denoted by  $f'_{ij}$ and $f''_{ij}$ respectively. Their odds-ratios and discrimination scores are $oz_1, oz_2, \delta_1, \delta_2$ respectively. We want to prove that when $D$ and $P$ are less correlated in $e_1$ than in $e_2$ (i.e., $|oz_1-1|<|oz_2-1|$), the discrimination in $e_1$ can be larger than that in $e_2$ (i.e., $|\delta_1|>|\delta_2|$). 

We consider only the case of $oz_1-1>0 \land oz_2-1>0$. Let $dz:=|oz_1-1|-|oz_2-1|=\frac{f'_{11}*f'_{00}}{f'_{10}*f'_{01}}-\frac{f''_{11}*f''_{00}}{f''_{10}*f''_{01}}$, and  $d\delta:=|\delta_1|-|\delta_2|=|\frac{f'_{11}}{f'_{11}+f'_{01}}-\frac{f'_{10}}{f'_{10}+f'_{00}}|-|\frac{f''_{11}}{f''_{11}+f''_{01}}-\frac{f''_{10}}{f''_{10}+f''_{00}}|$.  When want to show that when $dz<0$, there are cases satisfying $d\delta>0$.

We present a tuple counts table below.  
Assume non-zero positive integers $K,m,w$ and the following counts tables. 
\[\left[ \begin{array}{ll}\multicolumn{2}{l}{e_1:}\\ f_{11}=mK  &  f_{10}=K \\  f_{01}=K  & f_{00}=K \end{array}\right] \qquad 
  \left[\begin{array}{ll} \multicolumn{2}{l}{e_2:}\\ f'_{11}=K  &  f'_{10}=wK \\  f'_{01}=K  & f'_{00}=K \end{array} \right]         \] 
$dz=oz_1-oz_2=m-\frac{1}{w}$. $dz<0$  if $m<\frac{1}{w}$ plus $m>1\land w<1$ reflecting our assumption $oz_1>1\land oz_2>1$. On the other hand, $d\delta=|\delta_1|-|\delta_2|=|\frac{m}{m+1}-\frac{1}{1+1}|-|\frac{1}{1+1}-\frac{w}{w+1}|=because (m>0 \land w>0)=\frac{m}{m+1}-\frac{w}{w+1}$. $d\delta>0$ if $m>w$. Combining the conditions for $dz<0$ and $d\delta>0$, we have $\frac{1}{w}>m>max(w,1)$. For example, if we let $m=2$ and $w=0.2$, $dz=-3<0$ and $d\delta=2/3-0.2/1.2=1/2>0$. We proved that a data set having lower correlation (odds-ratio closer to 1) may have higher discrimination compared to another data set.

The item is approved.

\end{enumerate}
$\Box$

This result is very important to understanding the complexity between redline attributes and the discrimination of predictions. 
Let $A$ be a redline attribute. It is an intuition that higher correlation between $A$ and $P$ and higher correlation between $P$ and $D$ would lead to higher model discrimination. In a previous section we showed that the discrimination caused by redline attributes is modified by the error distribution of the model. Now we show that higher $corr(P,D)$ does not mean higher discrimination. Consequently the relationship between model discrimination and redline attributes is not intuitive.

\section {Exploration of discrimination levels in real data sets} 
In this section, we present the results of our exploration of discrimination in four real world data sets. We present the following results. (1) Data from real application is often discriminatory and how the number of explanatory attributes affect the score. (2) Classification models may change the level of discrimination in the predictions compared with the one in the train data. (3) Classification models built from non-discriminatory data may still be discriminatory.

\subsection{data sets}\label{sec:data set}
We use four real world data sets as shown in the following list. Name, size, source, and attributes of the data sets are shown. All data sets are processed to have binary (0,1) values. The values of ordinal attributes are binary-zed using median. Categorical attributes are binary-zed by taking majority and the rest.  The labels (P), (E), and (D) against some attributes indicate the types protected, explanatory, and outcome respectively. The attributes without a label are O-attributes.   

\begin{itemize}\small
             \topsep=1mm  \parsep=0mm \itemsep=0mm \parskip=0cm \itemindent=1em
\item[{\bf Adult}] US Census 1994. numb(rows)=48842; minority class rate=.25\\
     https://archive.ics.uci.edu/ml/data sets/adult \\
    Attributes: age45(P), natCountryUS(P), raceBlack(P), sexM(P), workPrivate(E), occuProf(E), workhour30(E), eduUni(E), relaNoFamily, married, income50K(D)

\item[{\bf Cana}] Canada Census 2011 \cite{canada-census} \footnote{The author wishes to acknowledge the statistical office that provided the underlying data making this research possible: Statistics Canada}. numb(rows)=691788; minority class rate=.34 \\
     https://international.ipums.org \\
      Attributes: weight100(P), age50(P), sexM(P), edUni(E), occProf(E)\{\}, occSkilled(E), occOther(E), hoursfull(E), govJob, classSalary, income45K(D)
      
\item[{\bf Germ}] German Credit. numb(rows)=1000; minority class rate=.3 \\
         https://archive.ics.uci.edu/ml/data sets/statlog+(german+credit+data) \\
         Attributes: age35(P), single(P), foreign(P), chkAccBal(E), duration20m(E), creditHistGood(E), purposeCar(E), credit2320(E), savings500(E), emp4y(E), installPct3(E), sexM(E), guarantor(E), resid3y(E), propertyYes(E), instPlanNon(E), houseOwn(E), creditAcc(E), jobSkilled(E), people2(E), hasTel(E), approved(D)

\item[{\bf Recid}]  Recidivate-violent \cite{larson2016-analyze-compas}; numb(rows)=4744; minority class rate=.14 \\
	https://github.com/propublica/compas-analysis\\
   Attributes: sexM(P), age30(P), raceAfrica(P), raceWhite(P), raceOther(P), juvFelonyCnt1(E), juvMisdCnt1(E), juvOthcnt1(E), priorsCnt3(E), cjail1Month(E), cChargeisdemM(E), isRecid(E), score8(D) \\
\end{itemize}
The Recidivism data set follows \cite{larson2016-analyze-compas}. The score8 column stores predictions from a system called COMPAS. The isRecid column stores whether the person re-committed a crime. We want to see if score8 values can be accurately re-predicted. 
 
 \vspace*{0.5ex}
We now present the results of experimental evaluation of our method.

\subsection{Discrimination in original data sets}
In the experiments, each (original) data set is stratified into E-groups by using the explanatory variables specified in the data set descriptions above. A discrimination score is calculated for each protected attribute in each E-group following Formula \bref{eq-count-dc}. The discrimination scores of different E-groups for the same protected attribute are averaged with the weights of the group sizes. The global discrimination score of the data set is the maximum of the averaged scores over different protected attributes. 

The results are shown in Table \ref{plot:dsc-orig-data} where $glbds$ is the global discrimination score. The table also lists the worst (maximal) E-group discrimination score ($wgds$) and the percentage ($wg\%$) of tuples in this worst group out of all the tuples in the data set. In the data set, the discrimination score of some groups is over the threshold $\alpha=0.05$ and these groups are called the over-limit groups. The discrimination scores of these groups are averaged to get the average score $ogds$ and the percentage of tuples in these groups is $og\%$. The top three protected attributes ranked by absolute values of the their score are listed in the right-most column.  

\begin{table}[h]
\center      
\includegraphics[scale=0.9]{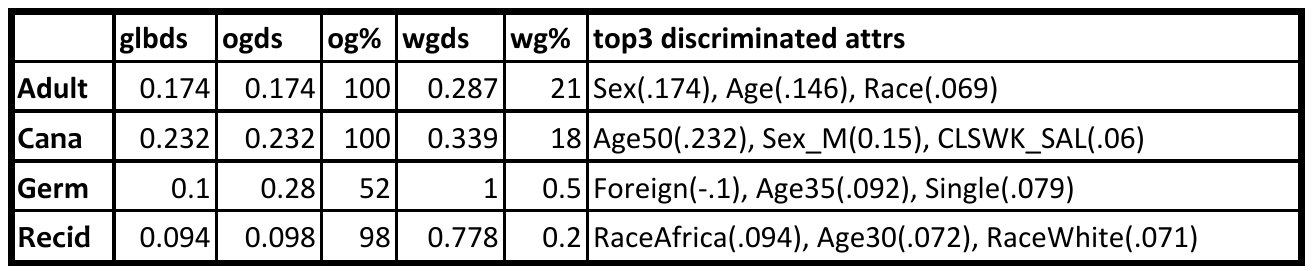}
\caption{Discrimination of original data sets}
\label{plot:dsc-orig-data}
\end{table}

For the Adult data, all E-groups are discriminatory (100\%) with the score 2 times more than the threshold, and the over-limit groups are 4 times more than the threshold.  The worst discrimination happened to the attributes of Gender and Age. 

In the Canada data, the discrimination scores are larger than those of the Adult data set. The worst discrimination happened to Age and Sex male. People who takes Salary (instead of Wage) were slightly discriminated. 

German Credit data's discrimination level is lower although some E-groups (with 52\% of tuples) have a discrimination score of 0.28, and the worst E-group (with 5 tuples or 0.5\%)  has a score of 1. After some investigation, we found that this extreme score is caused by a small group size. When the size of an E-group is small, the discrimination score can be dramatic. The tuple counts of this worst group is (3,0,0,2) and  the score calculation is $\frac{3}{3+0}-\frac{0}{0+2}=1$. 

The Recidivate data has an overall discrimination score of 0.098, lowest among all data sets. The worst discrimination (0.778) happened to the protected attribute RaceAfrica in an E-group with 11 tuples (0.2\% of the data set).  The tuple counts of the group is (7,2,0,2) and the score calculation is $\frac{7}{7+2}-\frac{0}{0+2}=0.778$. Among 9 Africans, 7 were predicted to recidivate, but among the 2 non-Africans, 0 were predicted to recidivate. 

We note that the above observations are conditional. They are dependent on the way in which data is discretized and on what and how many variables are specified as explanatory ones. 

\begin{figure}[h]
\center      
\includegraphics[scale=0.4]{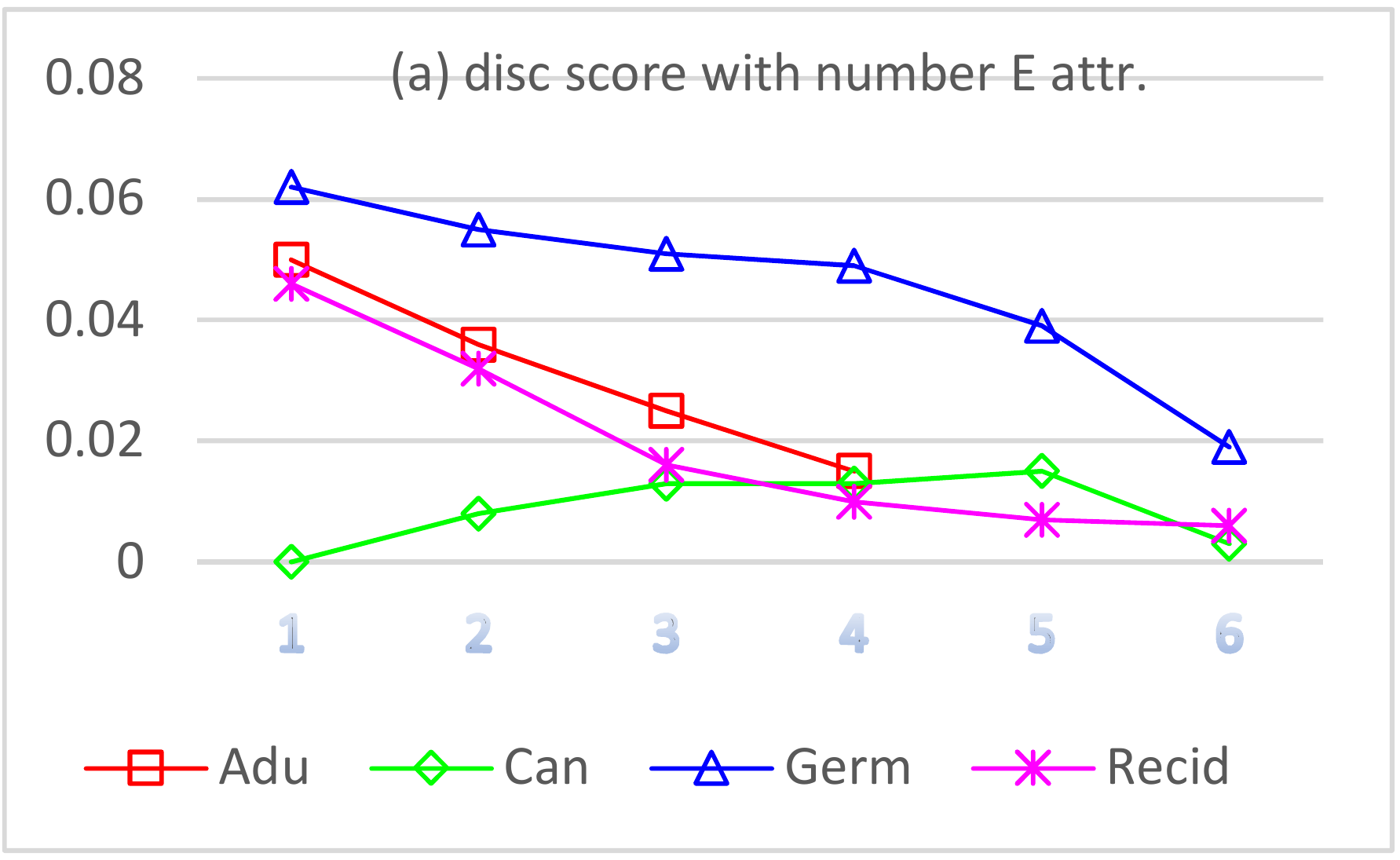}
\caption{Number of explanatory attributes on discrimination}
\label{plot:dsc-E-attr}
\end{figure}

Figure \ref{plot:dsc-E-attr} describes the relationship between discrimination and the number of explanatory attributes. The experiments are done by using a specified number of explanatory attributes among all the described explanatory variables in turn to calculate an average discrimination score. For example, with the Adult data set, there are 4 possible explanatory attributes in the description. In the case of using 1 explanatory attribute, we run 4 experiments, each with a different explanatory attribute and the scores from these 4 experiments are averaged to get the final score for the one explanatory variable case. From Figure \ref{plot:dsc-E-attr}, we observe that as the number of explanatory attributes increases, the discrimination score becomes lower. This trend is reasonable because when more explanatory attributes are used, more discrimination can be justified and consequently the discrimination level reduces. 

We note that as more explanatory attributes are used, the data becomes more fragmented and the sizes of strata of explanatory attributes reduce to small numbers or 0, which can make discrimination scores change dramatically to large values.

\subsection{Discrimination of classifiers}
In this exploration, we use commercially available modeling algorithms to build classifiers on the training data sets (the original data sets used above). The models are then used to predict a new outcome for each tuple in the training data sets. The predicted outcomes replace the original observed outcomes to form new data sets called predicted data sets. We calculate the discrimination scores on the predicted data sets. 

For the same training data set, we get different predicted data sets when the classifiers are different. We choose five well-used classification algorithms, namely decision tree (DT), Bayes network (BN), neural network (NN), logistic regression (LR), and support vector machine (SVM) from SAS Enterprise miner and use these algorithms with the default parameters to generate predicted data sets.  Discrimination scores are calculated for the predicted data sets and the results are shown in Table \ref{plot:dsc-eff-models}. BCR stands for balance classification rate which is the average of the true positive rate and the true negative rate. Err stands for misclassification rate.  These two measures indicate the quality of the classification models. Better models have larger BCR and smaller Err. 

\begin{table}[h]
\center      
\includegraphics[scale=0.8]{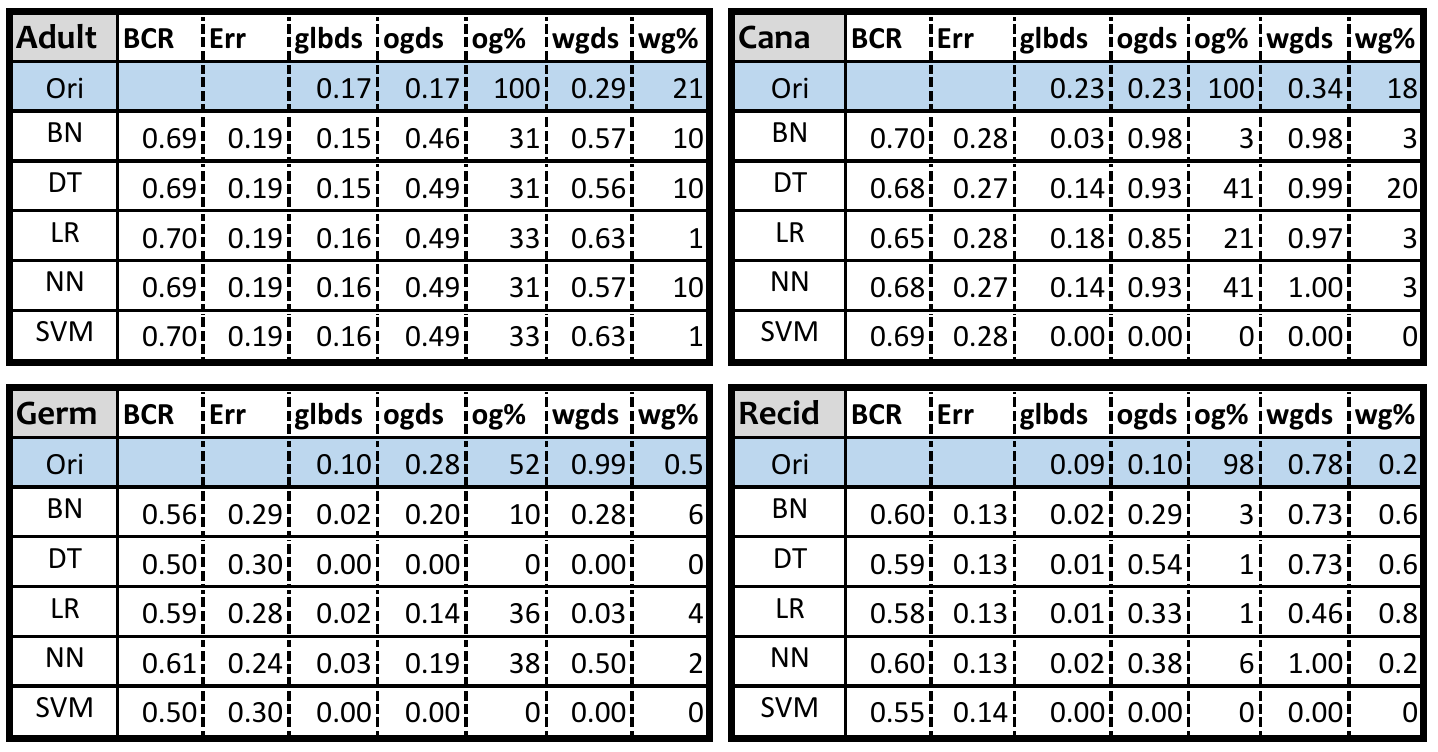}
\caption{Effect of models on discrimination of Adult data set}
\label{plot:dsc-eff-models}
\end{table}

First we look at the results for the Adult data set in the top-left corner. Compared to the discrimination scores of the original data (line Orig), the predicted data set has (1) a slightly lower global score (glbds), (2) much higher over-limit score (ogds) with a lower percentage (og\%), and (3) a much higher worst group score (wgds) with smaller percentage (wg\%). Points (2) and (3) indicate that the classification errors made by the models are quite high (19\% ), some of these errors made some E-groups less discriminatory, others made other groups worse. 

The Canada, German, and Recidivate data sets have similar properties. We note that the last line for SVM of these data sets have a global score of 0. The reason is not that the classifier is the best, but that the classifier made so many errors such that the error rate equals to the minor class rate in the data set. For example, in German credit data, the percentage of non-approved cases is 30\% while the method's misclassification rate is also 30\%. A close check found that the classifier predicted all negative class as positive class. In this case, all non-approved cases are predicted as approved cases. When all predictions have only one class, there is no discrimination. 

\emph{From this we see that when examining the discrimination of a classification model, we must consider the balanced accuracy and error rate. Otherwise, the conclusion may not be right.}

\subsection{Non-discriminatory data does not mean fair classifiers}

In this section, we use experiments to show the independence between discrimination and classifiers trained from the non-discriminatory data. 

In the experiments, our non-discriminatory data was generated from the CV method \cite{Calders2010}. More specifically, we use the original data $r$ as training data to run the CV implementation in \cite{Friedler18-testbed} by specifying Sex as the only protected attribute, rejecting other protected attributes described in Section \ref{sec:data set}.  The CV method produces a predicted data set which is close to non-discriminatory. We call the predicted data set  the \emph{$CV$ data set}. Further, we train a predictive classifier from the CV data set. The classifier again produces a predicted data set $r'$. The discrimination in $r'$ is what we want to analyze. That is, we want to see if the classifier from the non-discriminatory $CV$ data set is discriminatory. 

The results of this experiment are in Table \ref{plot:dsc-classifier-fairdata}. The grey line labelled by `CV' is the $CV$ data set. The data set $r'$ for each classifier is labelled by the classifier name. For example, $BN$ is the data set $r'$ from the Bayes network classifier.  Because the CV method does not support explanatory attributes, the results do not reflect group level discrimination. That is, the whole data set is seen as a large E-group for the only protected attribute Sex.

\begin{table}[h]
\center      
\includegraphics[scale=0.9]{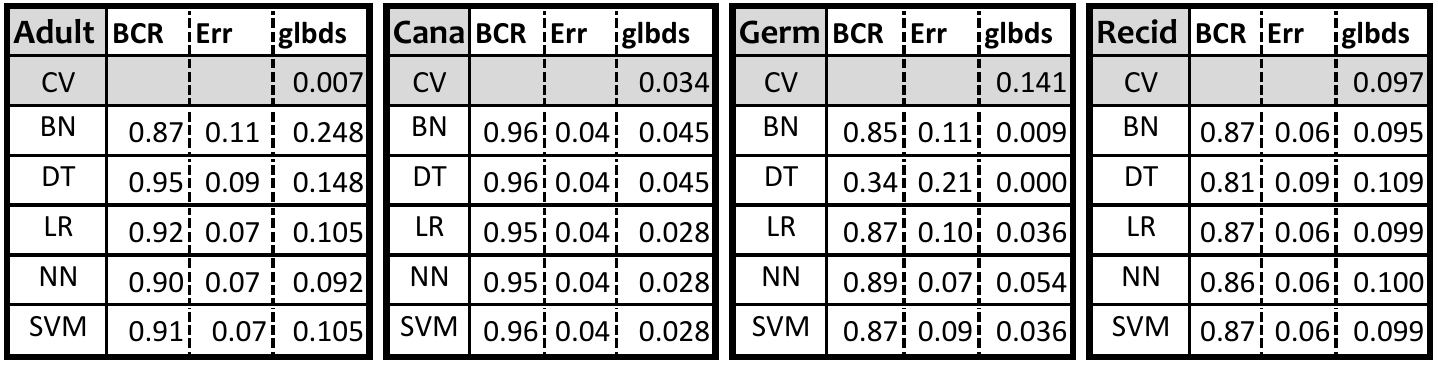}
\caption{Discrimination of classifiers trained on non-discriminatory data}
\label{plot:dsc-classifier-fairdata}
\end{table}

The results show that the discrimination of training data and the discrimination of the classifiers from the data are independent. 
Among the four data sets, Adult CV has the smallest discrimination score, but the classifiers from this data set have highest discrimination scores. Although different algorithms among BN, DT, LR, NN, and SVM have different balanced accuracy and different error rates, all of them are discriminatory and the scores are 13-33 times larger than the score on the CV data set.  

The results also show that high accuracy does not lead to lower discrimination. For example, DT for Adult CV data has highest BCR, but its discrimination score is not the least.  

The numbers for the German credit data set indicate the same independence but in an opposite way.  With this data set, the discrimination of the CV is over the threshold, but the discrimination of the classifiers is less than the threshold (NN is only slightly over). 

The numbers for the Canada and the Recidivate data sets show that the classifiers did not change the discrimination levels.  

From all these, we observe clearly that discrimination of classifiers are independent to the discrimination of training data. It is also independent of accuracy and error rate. These two conclusions are reasonable because with the same error rate, the error distribution in the protected attributes can be very different, leading different discrimination scores.

 \section{Related Work}
Algorithmic discrimination has attracted a lot of research effort. The work focused in two areas: discrimination detection and discrimination removal from data and from models. The work on removal are in three categories: pre-processing, manipulation of model learning algorithms and post processing. We now review the work done in these directions.
  
\smallpara{Discrimination detection in data} The core problem of detection is to define and choose metrics to measure discrimination. \citet{zlio-review-2017} has a good summary of previous metrics \cite{Pedreschi2008,Pedreschi2009,Zliobaite2011,Fukuchi2013,Ristanoski2013-imbalancedData,Ruggieri2014,Feldman2015,Fish2016}.  Some recently proposed causality-based metrics are \cite{Fish2016,Kleinberg2016,wu2016c,Zhang2016,Li-bigdata2017,ZhangLu2016dr} and the metrics for individual discrimination are \cite{Luong2011,Dwork2012,Mancuhan2014,Zhang2016}.

\smallpara{Removal of discrimination from training data (pre-processing)}  \citet{Feldman2015} proposed to transform data in a data set so that the red-line attributes, those that correlated to the target, become independent to the target variable. \citet{Friedler18-testbed} summarized some of the work in this direction.

\smallpara{Modification of model training algorithms} \citet{Kamiran2010} proposed to combine information gain and discrimination gain in {\it decision tree} learning and to use a post-relabelling process to remove discrimination from predictions.  \citet{Calders2010} adjusts the probability in naive Bayes methods so that the predictions are discrimination free.   
\citet{Kamishima2018-actual-indep} proposed a regularization method for logistic regression method.      \citet{Zafar2015}  represented discrimination constraints via a convex relaxation and optimized the accuracy and discrimination in the SVM learning algorithm. 
\citet{Woodworth2017} proposes a two step method to build a non-discriminatory classifier. The data set is divided into two subsets S1 and S2. In the first step, a classifier is built to minimize error rate under the constraints of discrimination in data set S1. In the second step, a post-processing model is built on data set S2.          
 \citet{edward18-continuous-T} proposed a discrimination-aware measure for decision tree induction for continuous data. 
\citet{landerio16aaai-undertraining} proposes to use a weight under-training method by strengthening confounder features to build  a model. 
 \citet{Kearns2018-subgroupfairness} used an optimization method in model learning.
\citet{Kamishima2018-actual-indep} has done a deep analysis of CV2NB, ROC and proposed a method called universal ROC. 

\smallpara{Removal of discrimination from predictions of a model (post-processing)}
The work of \citet{Kamiran2010} relabels the predictions of leaf nodes of a decision tree to achieve discrimination goal. The post-processing model of Step 2 in \cite{Woodworth2017} minimizes the discrimination using the target, the predicted target and the protected variables on the second half of the training data.   
   \citet{Hardt2016} uses equalized odds to build a model for post-prediction manipulation. 
   \citet{Kamishima2018-actual-indep} proposed a method called universal ROC.

\section{Conclusion}
In this paper, we analyzed some properties of algorithmic discrimination. The properties reveal relationships between discrimination and correlation, data partition, and classification models. These properties have important implications in building predictive models, assessing discrimination in  data sets and models, and evaluating discrimination-aware models. The paper also explored discrimination of real world data sets. The results of exploration show that discrimination does exist in real world data sets, and the discrimination of models is independent to the discrimination of training data. This implies that the fairness of models cannot be achieved by manipulating training data. Our future work is to develop a general method to achieve fairness independent of learning algorithms.

\bibliographystyle{ACM-Reference-Format}
\bibliography{refs} 

\end{document}